\documentclass[11pt,a4paper]{article}
\usepackage{jheppub}
\usepackage{setspace,braket}
\usepackage{amsmath, amssymb, bm}
\usepackage{etoolbox}

    \makeatletter
    \patchcmd{\maketitle}{\@fpheader}{}{}{}
\makeatother
\def\be{\begin{equation}}
\def\ee{\end{equation}}

\def\k{\kappa}

\def\La{\Lambda}

\def\ps{\psi}

\def\({\left(}
\def\){\right)}
\def\[{\left[}
\def\]{\right]}

\newcommand{\bea}{\begin{eqnarray}}
\newcommand{\eea}{\end{eqnarray}}

\def\d#1#2{\frac{\displaystyle #1}{\displaystyle #2}}
\def\no{\nonumber}

\numberwithin{equation}{section}
\renewcommand{\thesection}{\arabic{section}}

\begin{document}
\renewcommand{\thefootnote}{\fnsymbol{footnote}}

\title{Horizon thermodynamics in fourth-order gravity}
\author[a,b]{Meng-Sen Ma}
\affiliation[a]{Department of Physics, Shanxi Datong
University,  Datong 037009, China}
\affiliation[b]{Institute of Theoretical Physics, Shanxi Datong
University, Datong 037009, China}
\emailAdd{mengsenma@gmail.com; ms\_ma@sxdtdx.edu.cn}

\abstract{
In the framework of horizon thermodynamics, the field equations of Einstein gravity and some other second-order gravities can be rewritten as the thermodynamic identity: $dE=TdS-PdV$.
However, in order to construct the horizon thermodynamics in higher-order gravity, we have to simplify the field equations firstly. In this paper, we study the fourth-order gravity and convert it to second-order gravity via a so-called `` Legendre transformation " at the cost of introducing two other fields besides the metric field.
With this simplified theory, we implement the conventional procedure in the construction of the horizon thermodynamics in 3 and 4 dimensional spacetime. We find that the field equations in the fourth-order gravity can also be written as the thermodynamic identity. Moreover, we can use this approach to derive the same black hole mass as that by other methods.

}
\maketitle
\onehalfspace

%%%%%%%%%%%%%%%%%%%%%%%%%%%%%%%%%%%%%%%
%%%%%%%%%%%%%%%%%%%%%%%%%%%%%%%%%%%%%%%
%%%%%%%%%INTRODUCTION%%%%%%%%%%%%%%%%%%%
%%%%%%%%%%%%%%%%%%%%%%%%%%%%%%%%%%%%%%%
\renewcommand{\thefootnote}{\arabic{footnote}}
\setcounter{footnote}{0}
\section{Introduction}
\label{intro}

It has long been known that gravitational system has thermodynamic properties since the works of Hawking and Bekenstein\cite{Bekenstein-1973,Hawking-1975}.
Just like conventional thermodynamic systems, black holes also have the temperature, entropy and other thermodynamic quantities. Besides, black holes also have fruitful phase structures\cite{Hawking-1983,Chamblin-1999,
Chamblin-1999a,Peca-1999,Wu-2000,Dolan-2011,Banerjee-2012a,Gunasekaran-2012,Kubiznak-2012,Niu-2012,Hendi-2013,Zhao-2013,Dehghani-2014,Ma-2014}.
Not only that, it was found that the field equations of Einstein gravity and other more general gravitational theories, such as $f(R)$ gravity,  can be derived from an equation of state of local spacetime thermodynamics\cite{Jacobson-1995,Elizalde-2008}.

There is another route to explore the relationship between the gravitational system and its relevant thermodynamic properties. It is the framework of horizon thermodynamics proposed by Padmanabhan\cite{Padmanabhan-2002}.
It is shown that Einstein's field equations for a spherically symmetric spacetime can be written in the form of thermodynamic identity: $dE=TdS-PdV$. This makes the connection between gravity and thermodynamics more closely. The radial pressure $P$ is the $(^r_r)$ component of energy-momentum tensor. This approach has also been extended to the non-spherically symmetric cases\cite{Kothawala-2007,Akbar-2007}  and other theories of gravity, such as Lovelock gravity\cite{Paranjape-2006}, Ho\v{r}ava-Lifshitz theory\cite{Cai-2010} and Einstein gravity with conformal anomaly\cite{Son-2013}.

However, we can notice that many previous works on horizon thermodynamics were based on the second-order gravities. This means that in the field equations there are at most the second-order derivatives of metric functions. In fact, one can find that in many cases only the first-order derivatives of metric functions exist. The work \cite{Cai-2010} on the Ho\v{r}ava-Lifshitz theory is the first study on horizon thermodynamics in higher-order derivative gravity. But it is shown that the field equations of Ho\v{r}ava-Lifshitz gravity in the static, spherically symmetric case, only include the first-order derivatives of metric functions. Generally, in higher derivative gravities, the field equations are full of higher-order derivatives of metric functions and are very complicated. We cannot directly extend the previous approach to these theories.
We should first reduce the higher-derivative gravity to some lower-derivative gravity. This process can be done via a `` Legendre " transformation\cite{Magnano-1987,Magnano-1990,Jakubiec-1988}. One can even convert the higher-derivative gravity from the original Jordan frame into Einstein frame by a conformal transformation \cite{Whitt-1984,Maeda-1989} or field redefinition\cite{Jacobson-1994}. Under some conditions, one can verify the equivalence of black hole thermodynamics between the two frames\cite{Jacobson-1994,Koga-1998}. However, we do not want to deal with the horizon thermodynamics of higher-derivative gravity in the Einstein frame, because there is still no consensus on the physical equivalence between the Jordan frame and the Einstein frame\cite{Magnano-1994,Faraoni-1999,Capozziello-2010}. In this paper we will study a fourth-order derivative gravity. Via the `` Legendre " transformation, it can be reduced to a second-order derivative gravity with some additional auxiliary fields, which is still equivalent to the original fourth-order derivative gravity. In this way, the field equations can be simplified greatly. Thus, we can extract the useful information from the field equations to construct the horizon thermodynamics.

The plan of this paper is as follows:
In Sec.2 we give a very short introduction to horizon thermodynamics in Einstein gravity.
We present the necessary demonstrations on some notations.
In Sec.3 we introduce the fourth-derivative gravity theory and obtain the second-derivative gravity via the `` Legendre " transformation.
In Sec.4 we give some examples to show the horizon thermodynamics in fourth-derivative gravity in 3 and 4 dimensional spacetime.
In Sec.5 we summarize our results and discuss the possible future directions.
In Appendix, we give the complete form of some field equations in components.

\section{Horizon thermodynamics in Einstein gravity}

In this section, we simply introduce the horizon thermodynamics in Einstein gravity first proposed in \cite{Padmanabhan-2002}. For a static, spherically symmetric spacetime, the metric in the
Schwarzschild gauge can be written as
\be\label{metric4d}
ds^2=-f(r)dt^2+f(r)^{-1}dr^2 + r^2d\Omega^2.
\ee
Substituting the metric into  Einstein field equation
\be\label{efe}
G^\mu_{~\nu}=R^\mu_{~\nu}-\d{1}{2}R g^\mu_{~\nu}=8\pi T^\mu_{~\nu},
\ee
one can obtain
\be\label{ht1}
r_{+} f'(r_{+})-1=8\pi r_{+}^2P,
\ee
and thus
\be\label{1-law}
d\left(\d{r_{+}}{2}\right)=\d{f'(r_{+})}{4\pi}d(\pi r_{+}^2)-PdV,
\ee
where $r_{+}$ represents the position of the event horizon, which must satisfy $f(r_{+})=0$. $P=T^r_{~r}|_{r=r_{+}}$, is the radial pressure of matter at the horizon. $V=4\pi r_{+}^3/3$ is called the ``areal volume". According to Eq.(\ref{metric4d}), it is just the volume of the black hole with horizon radius $r_{+}$ in the coordinate.

Considering the temperature of the black hole is
\be\label{tmp}
T=\d{\kappa}{2\pi}=\d{f'(r_{+})}{4\pi},
\ee
Eq.(\ref{1-law}) is just the conventional thermodynamic identity
$dE=TdS-PdV$ with $E=r_{+}/2, ~ S=A/4=\pi r_{+}^2$. In the source-free case, the metric function represents Schwarzschild black hole. For this black hole, $E$ is just the mass $M$ of the black hole.

This result above only depends on the theories of gravity under consideration.
It has nothing to do with the concrete black hole solution. The contributions from matter fields have been contained in the pressure $P$. Obviously, except for vacuum cases,
$E\neq M$ generally.

In this case, only two pairs of thermodynamic variables exist, which are the intensive quantities $(T,~P)$ and the extensive quantities $(S,~V)$.
In this framework the thermodynamic properties are directly related to the gravitational theories under consideration.
The details of matter content are not important and the concrete black hole solutions are also not necessary.
In this framework, we have studied the phase transitions and thermodynamic stabilities of black holes in general relativity and Gauss-Bonnet gravity\cite{Ma-2015a}.

\section{The fourth-order gravity}

In this section we will generalize the original horizon thermodynamics approach to the higher-derivative gravity.
Let us consider a fourth-derivative gravity action
\be\label{action}
S=S_G +S_{M}=\int d^{d}x \left(\mathcal{L}_G+\mathcal{L}_M\right)=\int d^{d}x \sqrt{-g}\left(L_G+L_M\right),
\ee
where $S_{M}$ represents the action of matter fields, and the gravitational Lagrangian
$L_G$ takes the form
\be\label{LG}
L_G=\frac{1}{\kappa}\left(R-2\Lambda+\alpha R^2+\beta R_{\mu\nu}R^{\mu\nu}\right),
\ee
with $\kappa=16\pi G_d$. It should be noted that the matter fields are necessary to derive the horizon thermodynamics, although its concrete form is not necessary. We need energy-momentum tensor to determine the $PdV$ term uniquely.

The field equations that follow from the action Eq. (\ref{action}) are
\be\label{feq}
\mathcal {G}_{\mu\nu}+E_{\mu\nu}=8\pi G_d T_{\mu\nu},
\ee
%where the energy-momentum tensor takes the standard definition:
%\be
%T_{\mu\nu}=-\frac{2}{\sqrt{-g}}\frac{\delta S_M}{\delta g^{\mu\nu}},
%\ee
where
\bea
{G}_{\mu\nu}&=&R_{\mu\nu}-\frac{1}{2}g_{\mu\nu}R+\Lambda g_{\mu\nu}, \no \\
E_{\mu\nu} &=& 2\beta(R_{\mu\rho}\, R_\nu{}^\rho - R^{\rho\sigma}
  R_{\rho\sigma}\, g_{\mu\nu}) + 2\alpha R\, (R_{\mu\nu} - R\, g_{\mu\nu})\no \\
  && +\beta\, ( \square R_{\mu\nu} + \nabla_\rho\nabla_\sigma R^{\rho\sigma}\,
  g_{\mu\nu} - 2\nabla_\rho \nabla_{(\mu} R_{\nu)}{}^\rho) +
 2\alpha\, (g_{\mu\nu}\, \square R -\nabla_\mu\nabla_\nu R).
\eea
If substituting the metric (\ref{metric4d}) into these field equations, the expressions are so complicated that one cannot directly construct the horizon thermodynamics.

Now we employ the `` Legendre " transformation to simplify the field equations. We can introduce two conjugate fields in the following way:
\be
\Phi=\d{\delta \mathcal{L}}{\delta R}, \quad \Psi^{\mu\nu}=\d{\delta \mathcal{L}}{\delta R_{\mu\nu}},
\ee
where $\mathcal{L}=\mathcal{L}_G+\mathcal{L}_M$.

We can further set $\sqrt{-g}\phi=\Phi$ and $\sqrt{-g}\psi^{\mu\nu}=\Psi^{\mu\nu}$. In this way, we can obtain
\be\label{conjugate}
\phi=\d{1}{\kappa}(1+2\alpha R), \quad \psi^{\mu\nu}=\d{2\beta}{\kappa}R^{\mu\nu}.
\ee
Now we take the `` Legendre " transformation according to the two pairs of conjugated quantities. First, we should invert Eq. (\ref{conjugate}) to obtain $R$ and $R_{\mu\nu}$ as functions of $\phi$ and $\psi_{\mu\nu}$, respectively.
This can be easily done. Then substituting them into the following definition:
\be
\mathcal{H}(\phi,\psi_{\mu\nu})=\Phi R+\Psi^{\mu\nu}R_{\mu\nu}-\mathcal{L}=\sqrt{-g}\left[\d{(\kappa \phi-1)^2}{4\alpha\kappa}+\d{\kappa}{4\beta}\psi^{\mu\nu}\psi_{\mu\nu}+\d{2\La}{\kappa}-L_M\right].
\ee
At last, we define
\bea\label{LH}
\mathcal{L}_H(g_{\mu\nu},\phi,\psi_{\mu\nu})&=&\Phi R+\Psi^{\mu\nu}R_{\mu\nu}-\mathcal{H}(\phi,\psi_{\mu\nu})\no \\
&=&\sqrt{-g}\left[\phi R+\psi^{\mu\nu}R_{\mu\nu}-\d{2\La}{\kappa}-\d{(\kappa\phi-1)^2}{4\alpha\kappa}-\d{\kappa}{4\beta}\psi^{\mu\nu}\psi_{\mu\nu}+L_M\right].
\eea
Treating $g_{\mu\nu},\phi,\psi_{\mu\nu}$ as three independent field variables, the variation of the Lagrangian (\ref{LH}) yields the following field equations:
\bea
0&=&\d{\delta \mathcal{L}_H}{\delta \phi}= R +\d{1-\kappa\phi}{2\alpha},\label{eq1} \\
0&=&\d{\delta \mathcal{L}_H}{\delta \psi^{\mu\nu}}=R_{\mu\nu}-\d{\k}{2\beta}\ps_{\mu\nu},\label{eq2} \\
0&=&\d{\delta \mathcal{L}_H}{\delta g^{\mu\nu}}=\d{g_{\mu\nu}}{8\alpha \k}+\d{\La g_{\mu\nu}}{\k}+\d{\k }{2\beta}\psi_{\mu}^{~\rho}\psi_{\nu\rho}+\d{\k }{8\beta}g_{\mu\nu}\psi_{\rho\sigma}\psi^{\rho\sigma}-\d{1}{2}g_{\mu\nu}\psi^{\rho\sigma}R_{\rho\sigma}-\d{g_{\mu\nu}\phi}{4\alpha}\no \\
&-&\d{1}{2}g_{\mu\nu}R\phi+\d{\k}{8\alpha}g_{\mu\nu}\phi^2-\nabla_\mu\nabla_\nu \phi-\nabla_{\rho}\nabla_{(\mu}\psi_{\nu)}^{~\rho}+\d{1}{2}\Box \psi_{\mu\nu}+g_{\mu\nu}\Box\phi+\d{1}{2}g_{\mu\nu}\nabla_{\rho}\nabla_{\sigma}\psi^{\rho\sigma}\no \\
&-& \frac{1}{2}T_{\mu\nu}.\label{eq3}
\eea

In this way, the original fourth-order differential equations for $g_{\mu\nu}$ are reduced to several second-order differential equations at the price of introducing two other fields $(\phi,\psi_{\mu\nu})$.
Clearly, Eqs.(\ref{eq1}),(\ref{eq2}) are just Eq.(\ref{conjugate}). When substituting them into Eq.(\ref{eq3}), we again obtain the original fourth-order equation (\ref{feq}).

In the following, we shall consider some examples with $d=3,4$. In higher dimensional spacetime ($d\geq 5$), the gravitational Lagrangian (\ref{LG}) is no longer general. The term quadratic in Riemann tensor $R_{\mu\nu\alpha\beta}$ should also be included. This generalization is direct but nontrivial. One needs to introduce another field conjugated to $R_{\mu\nu\alpha\beta}$. However in this paper we will not consider this case.

\section{Horizon thermodynamics in fourth-order gravity}

\subsection{3D cases}

We take the metric ansatz
\be\label{metric3s}
ds^2=-f(r)dt^2+f(r)^{-1}dr^2 + r^2d\varphi^2.
\ee
With this metric ansatz one can easily derive Ricci tensor:
\be
R^t_{~t}=R^r_{~r}=-\frac{r f''(r)+f'(r)}{2 r}, \quad R^\varphi_{~\varphi}=-\frac{f'(r)}{r}.
\ee
Thus, according to Eq.(\ref{eq2}) $\psi^{\mu}_{~\nu}$ has the same symmetries as Ricci tensor. we can set
\be
\psi^t_{~t}= \psi^r_{~r}=p(r), \quad \psi^\varphi_{~\varphi}=h(r),
\ee
for short.
With these field variables, we can obtain the component equations of Eq.(\ref{eq3}). The complete forms are very lengthy and we put them in the Appendix. It is easy to see that the components $(^t_t)$ and $(^r_r)$ are not equal in general. However, the two components are the same at the horizons where $f(r_{+})=0$. Provided that the energy-momentum tensor of matter fields satisfy $T^t_{~t}=T^r_{~r}$, the $(^r_r)$ component equation will be
\bea
 &&r_+ \left(8 \alpha  \beta  \Lambda +\beta +4 \alpha  \beta  \kappa  p f''+4 \alpha  \beta  \kappa  f' p'+4 \alpha  \beta  \kappa  f' \phi '+\alpha  \kappa ^2 h{}^2+6 \alpha  \kappa ^2 p{}^2+\beta  \kappa ^2 \phi {}^2-2 \beta  \kappa  \phi \right) \no \\
&& + 4 \alpha  \beta  \kappa  f' \left(2 p+\phi \right)=4 \alpha  \beta  \kappa  r_+ T^r_{~r},
\eea
where the `` $'$ " represents derivative with respect to $r_{+}$.

According to Wald's formula\cite{Wald-1993,Iyer-1994},
\be
S=-2\pi\oint _\Sigma \d{\delta L_G}{\delta R_{abcd}}\hat \epsilon_{ab}\hat\epsilon_{cd}\bar\epsilon,
\ee
we can directly calculate the Noether charge entropy of black holes based on the gravitational Lagrangian (\ref{LG}),
\be
S=\d{4\pi A_H}{\kappa}(1+2\alpha R+2\beta R^r_{~r})=4\pi A_H(\phi+ p)=8\pi^2 r_{+}(\phi+ p),
\ee
where the cross-section area $A_H=2\pi r_{+}$ in 3D case.
We also know the temperature of black holes with the metric (\ref{metric3s}) is given by
\be
T=\d{f'(r_{+})}{4\pi}.
\ee
According to the temperature and the entropy, we arrange the field equation to a new form
\bea
&&4 \alpha  \beta  \kappa  r_+ T^r_{~r}=f' \left[4 \alpha  \beta  \kappa  r_+ \left(p'+\phi '\right)+4 \alpha  \beta  \kappa  p+4 \alpha  \beta  \kappa  \phi \right] \no \\
&+&4 \alpha  \beta  \kappa  r_+ p f''+4\alpha  \beta  \kappa  p f'+\alpha  \kappa ^2 r_+ h{}^2+6 \alpha  \kappa ^2 r_+ p{}^2+8 \alpha  \beta  \Lambda  r_++\beta  \kappa ^2 r_+ \phi {}^2-2 \beta  \kappa  r_+ \phi +\beta  r_+
\eea
Multiply the equation on the both sides with a constant $\d{\pi}{2\alpha\beta\kappa}$, we can obtain
\bea
2\pi r_{+} P&=&\d{f'}{4\pi} 8\pi^2 \left[r_{+}(\phi+p)\right]'\no \\
            &-&\left[\frac{\pi\left(2 \kappa r_+ \phi- r_+ - \kappa^2  r_+ \phi {}^2 -8\alpha  \Lambda  r_+\right) }{2\alpha\kappa }-\frac{\pi  \left(\kappa  r_+ h{}^2+6 \kappa  r_+ p{}^2\right)}{2 \beta }-2 \pi p  (r_+ f')'\right],
\eea
where we have taken $T^r_{~r}=P$.
Multiply $dr_{+}$ on the both sides, one can see that the above equation can be written into
\bea \label{3dpdv}
\left[\frac{\pi\left(2 \kappa r_+ \phi- r_+ - \kappa^2  r_+ \phi {}^2 -8\alpha  \Lambda  r_+\right) }{2\alpha\kappa }-\frac{\pi  \left(\kappa  r_+ h{}^2+6 \kappa  r_+ p{}^2\right)}{2 \beta }-2 \pi p  (r_+ f')'\right]dr_{+}=TdS-PdV
\eea
where $V=\pi r_{+}^2$ for 3D spacetime.

If the field equation can be written into the thermodynamic identity: $dE=TdS-PdV$, we guess that the term on the LHS in Eq.(\ref{3dpdv}) should correspond to $dE$. Especially, in the source-free case, $E=M$, there should be
\bea\label{mass3d}
M&=&\int \left[\frac{\pi\left(2 \kappa r_+ \phi- r_+ - \kappa^2  r_+ \phi {}^2 -8\alpha  \Lambda  r_+\right) }{2\alpha\kappa }-\frac{\pi  \left(\kappa  r_+ h{}^2+6 \kappa  r_+ p{}^2\right)}{2 \beta }-2 \pi p  (r_+ f')'\right]dr_{+}\no \\
&=& -\d{2\pi}{\kappa}\int \left[r_{+}(2\Lambda+\alpha R^2)+r_{+}\beta((R^\varphi_{~\varphi})^2+6 (R^r_{~r})^2)+2\beta R^r_{~r}(r_{+}f')'\right] dr_{+} .
\eea
Now let us check Eq.(\ref{mass3d}). When $\alpha=\beta=0$, the gravitational theory returns to Einstein gravity. There is the non-rotating BTZ black hole solution in the form of (\ref{metric3s}) with
\be\label{BTZ}
f(r)=-8G_3M+\d{r^2}{l^2},
\ee
where $\Lambda=-\d{1}{l^2}$. Substituting the metric function into Eq.(\ref{mass3d}), one can derive
\be
M_{GR}=\d{r_{+}^2}{8G_3l^2},
\ee
which is just the mass $M$ of BTZ black hole.

BTZ black hole also exists in the new massive gravity (NMG)\cite{Bergshoeff-2009}, in the notation of Eq.(\ref{LG}) which corresponds to the coefficients
\be
\alpha=\frac{3}{8m^2}\, , \qquad \beta=-\frac{1}{m^2}.
\ee
The non-rotating BTZ black hole solution also has the form of Eq.(\ref{BTZ}). However, due to the existence of correction term in the NMG, in this case the cosmological constant $\Lambda$ satisfies a slightly complicated relation:
\be
\Lambda+\d{1}{l^2}+\d{1}{4m^2l^4}=0.
\ee
Clearly, the conventional relation $\Lambda=-\d{1}{l^2}$ will recover in the General Relativity limit $m^2 \rightarrow \pm \infty$.

We can derive the mass of the black hole
\be
M_{NMG}=\d{2m^2l^2r_{+}^2-1}{16 G_3 m^2l^4}=\left(1-\d{1}{2m^2l^2}\right)M_{GR}.
\ee
One can also compute the black hole mass according to Abbot-Deser-Tekin (ADT) approach\cite{Abbott-1982,Deser-2002,Deser-2003}. It is shown that the result is the same \cite{Myung-2014}.

\subsection{4D cases}

In this case, we take the metric ansatz (\ref{metric4d}). Ricci tensor satisfies the relation:
\be
R^t_{~t}=R^r_{~r}=-\frac{f''(r)}{2}-\frac{f'(r)}{r}, \quad R^\theta_{~\theta}=R^\varphi_{~\varphi}=-\frac{r f'(r)+f(r)-1}{r^2}.
\ee
Also we set
\be
\psi^t_{~t}= \psi^r_{~r}=p(r), \quad \psi^\theta_{~\theta}=\psi^\varphi_{~\varphi}=h(r).
\ee
In this case, the temperature and entropy have the same form as the 3D case. However, the cross-section area $A_{H}$ is different in different spacetime dimension. In the 4D case, $A_{H}=4\pi r_{+}^2$. Thus the entropy is
\be
S=4\pi A_H(\phi+ p)=16\pi^2r_{+}^2(\phi+p).
\ee
Substituting the metric (\ref{metric4d}) into the field equation Eq.(\ref{eq3}), we can obtain four component equations, in which the $(^t_t)$ and $(^r_r)$ components are not equal as usual, but the
$(^\theta_\theta)$ and $(^\varphi_\varphi)$ components are the same. The complete expressions of $(^t_t)$ and $(^r_r)$ components are also given in the Appendix. Similar to the 3D case, when considering the equations at the horizon where $f(r_{+})=0$, the two equations are the same again. It is
\bea
&&\left(\frac{1}{\alpha  \kappa }+4 p f''+\frac{2 \kappa  h{}^2}{\beta }-\frac{8 h}{r_+^2}+\frac{8 \Lambda }{\kappa }+\frac{6 \kappa  p{}^2}{\beta }+\frac{\kappa  \phi {}^2}{\alpha }-\frac{2 \phi }{\alpha }-\frac{8 \phi }{r_+^2}\right) \no \\
 &+& f' \left(4 p'+\frac{16 p}{r_+}+4 \phi '+\frac{8 \phi }{r_+}\right)=4T^r_{~r}.
\eea
Considering  the temperature and the entropy and multiplying $\pi r_{+}^2dr_{+}$, we can rewrite the equation into the following form:
\bea
PdV=TdS+\left(\frac{1}{\alpha  \kappa }+4 p f''+\frac{2 \kappa  h{}^2}{\beta }-\frac{8 h}{r_+^2}+\frac{8 \Lambda }{\kappa }+\frac{6 \kappa  p{}^2}{\beta }+\frac{\kappa  \phi {}^2}{\alpha }-\frac{2 \phi }{\alpha }-\frac{8 \phi }{r_+^2}+\d{8pf'}{r_{+}}\right)\pi r_{+}^2dr_{+}.
\eea
Therefore, we also guess the mass of the black hole should be
\bea\label{mass4d}
M &=&-\int \left(\frac{1}{\alpha  \kappa }+4 p f''+\frac{2 \kappa  h{}^2}{\beta }-\frac{8 h}{r_+^2}+\frac{8 \Lambda }{\kappa }+\frac{6 \kappa  p{}^2}{\beta }+\frac{\kappa  \phi {}^2}{\alpha }-\frac{2 \phi }{\alpha }-\frac{8 \phi }{r_+^2}+\d{8pf'}{r_{+}}\right)\pi r_{+}^2dr_{+}\no \\
&=& -\d{4\pi}{\kappa} \int \left[ { r_+^2 \left(6 (R^r_{~r})^2 \beta +2 R^r_{~r} \beta  f''+2 \beta  (R^\theta_{~\theta})^2+2 \Lambda +\alpha  R^2\right)+4 R^r_{~r} \beta  r_+ f'}\right. \no \\
&-& \left.{2 (2 \beta  R^\theta_{~\theta}+2 \alpha  R+1)}\right]dr_{+}.
\eea
In Einstein gravity with cosmological constant, the black hole solution based on the metric (\ref{metric4d}) is Schwarzschild-(A)dS spacetime
\be
f(r)=1-\d{2G_4M}{r}-\d{\Lambda r^2}{3}.
\ee
From Eq.(\ref{mass4d}) one can easily compute the mass of the black hole
\be
M_{GR}=\d{1}{2G_4}\left(r_{+}-\d{\Lambda r_{+}^3}{3}\right)=M.
\ee
We also know that Schwarzschild-(A)dS black hole also exists in the gravitational theory (\ref{LG}).
According to Eq.(\ref{mass4d}), we can calculate the mass
\be
\mathcal{M}=\left[1+2\Lambda(\beta+4\alpha)\right]M,
\ee
This result is also consistent with ADT approach. For the so-called critical gravity \cite{Lue-2011} in which the coefficients satisfy
$\beta=-3\alpha=3/2\Lambda$, the black hole mass is clearly zero.

\section{Conclusion and Discussion}
\label{Conclusions}

In this paper we derived the horizon thermodynamics of spherically symmetric
black holes in a kind of fourth-order derivative gravity . The key idea of horizon thermodynamics approach is to write the field equations into the thermodynamic identity: $dE=TdS-PdV$. Because the original field equations
 of the fourth-order derivative gravity are very complicated. We are compelled to seek for simplification. Fortunately, the fourth-order derivative gravity can be reduced to
 a second-order derivative gravity via the so-called ``Legendre transformation ". Although some other fields are introduced, the metrics are left unchanged. So we can construct the horizon thermodynamics in the
second-order derivative gravity. Not like the previous works on horizon thermodynamics in second-order derivative gravity, where one can derive the black hole mass and entropy all together and often do not use the explicit black hole solutions, in fourth-order derivative gravity  one needs the concrete black hole solution to derive the black hole mass in general.

To avoid even more cumbersome computations, we focused on the static spherically symmetric case with $g_{00}=-1/g_{11}$. It would be plausible to extend our current study to a slightly more general case with $g_{00} \neq -1/g_{11}$. Another interesting future study would be to consider similar
investigations for black holes with more general gravitational Lagrangian, such as including the term quadratic in Riemann tensor or Weyl tensor.
It is also interesting to consider the horizon thermodynamics in higher-order gravity in Einstein frame, which may further shed light on the relations between the Jordan frame and the Einstein frame.

\appendix

\section{Complete forms of component field equations}

In this section we will give the complete form of the component field equations.

3D case:
\bea
(^t_t): &&\frac{1}{8 \alpha  \kappa }+\frac{1}{2} p(r) f''(r)+\frac{1}{2} f'(r) p'(r)+\frac{p(r) f'(r)}{r}+\frac{1}{2} f'(r) \phi '(r)+\frac{\phi (r) f'(r)}{2 r}-\frac{f(r) h'(r)}{2 r} \no \\
        &+&f(r) p''(r)+\frac{3 f(r) p'(r)}{2 r}+f(r) \phi ''(r)+\frac{f(r) \phi '(r)}{r}+\frac{\kappa  h(r)^2}{8 \beta }+\frac{\Lambda }{\kappa }+\frac{3 \kappa  p(r)^2}{4 \beta }+\frac{\kappa  \phi (r)^2}{8 \alpha }\no \\
        &-&\frac{\phi (r)}{4 \alpha }=\d{1}{2} T^t_{~t}
\eea
\bea
(^r_r):&&\frac{1}{8 \alpha  \kappa }+\frac{1}{2} p(r) f''(r)+\frac{1}{2} f'(r) p'(r)+\frac{p(r) f'(r)}{r}+\frac{1}{2} f'(r) \phi '(r)+\frac{\phi (r) f'(r)}{2 r}+\frac{f(r) h'(r)}{2 r}\no \\
       &+&\frac{f(r) p'(r)}{2 r}+\frac{f(r) \phi '(r)}{r}+\frac{\kappa  h(r)^2}{8 \beta }+\frac{\Lambda }{\kappa }+\frac{3 \kappa  p(r)^2}{4 \beta }+\frac{\kappa  \phi (r)^2}{8 \alpha }-\frac{\phi (r)}{4 \alpha }=\d{1}{2} T^r_{~r}
\eea

4D case:
\bea
(^t_t): &&\frac{8 f(r) \left(r \left(-h'(r)+r p''(r)+3 p'(r)+r \phi ''(r)+2 \phi '(r)\right)+p(r)+\phi (r)\right)}{r^2}+\frac{1}{\alpha  \kappa }+4 p(r) f''(r)\no \\
        &+&4 f'(r) p'(r)+\frac{16 p(r) f'(r)}{r}+4 f'(r) \phi '(r)+\frac{8 \phi (r) f'(r)}{r}-\frac{8 h(r)}{r^2}+\frac{2 \kappa  h(r)^2}{\beta }+\frac{8 \Lambda }{\kappa }\no \\
        &+&\frac{6 \kappa  p(r)^2}{\beta }-\frac{8 \phi (r)}{r^2}+\frac{\kappa  \phi (r)^2}{\alpha }-\frac{2 \phi (r)}{\alpha }=4 T^t_{~t}
\eea
\bea
(^r_r):&&\frac{8 f(r) \left(r \left(h'(r)+p'(r)+2 \phi '(r)\right)+p(r)+\phi (r)\right)}{r^2}+\frac{1}{\alpha  \kappa }+4 p(r) f''(r)+4 f'(r) p'(r)\no \\
  &+&\frac{16 p(r) f'(r)}{r}+4 f'(r) \phi '(r)+\frac{8 \phi (r) f'(r)}{r}-\frac{8 h(r)}{r^2}+\frac{2 \kappa  h(r)^2}{\beta }+\frac{8 \Lambda }{\kappa }+\frac{6 \kappa  p(r)^2}{\beta }\no \\
  &-&\frac{8 \phi (r)}{r^2}+\frac{\kappa  \phi (r)^2}{\alpha }-\frac{2 \phi (r)}{\alpha }=4 T^r_{~r}
\eea

\acknowledgments
The author will thank the anonymous referee for providing
constructive comments. I would  like to thank Prof. Ren Zhao for illuminating conversations.
This work is supported in part by the National Natural Science Foundation
of China (Grant Nos.11605107, 11475108) and by the Doctoral Sustentation Fund of Shanxi Datong
University (2011-B-03).

\bibliographystyle{JHEP}
\bibliography{H:/mms/References/references}

\end{document}